\documentclass[prl, showpacs, twocolumn, superscriptaddress]{revtex4}
\usepackage{graphics}

\newcommand{\bra}{\left\langle}
\newcommand{\ket}{\right\rangle}

\newcommand{\der}[2]{\frac{d #1}{d  #2}}

\newcommand{\pder}[2]{\frac{\partial #1}{\partial  #2}}

\newcommand{\fder}[2]{\frac{\delta #1}{\delta #2}}

\newcommand{\D}{\mathcal{D}}

\renewcommand{\vec}[1]{\mbox{\boldmath $#1$}}
\newcommand{\svec}[1]{\mbox{{\scriptsize \boldmath $#1$}}}
\newcommand{\tens}[1]{\mbox{\boldmath $\mathsf{#1}$}}

\newcommand{\tr}{\mathrm{tr}}

\newcommand{\ve}{\varepsilon}
\newcommand{\F}{\mathcal{H}}
\newcommand{\vphi}{\varphi}
\newcommand{\V}{\mathcal{V}}
\newcommand{\W}{\mbox{\boldmath $\mathcal{W}$}}
\newcommand{\Ws}{\mbox{\scriptsize \boldmath $\mathcal{W}$}}

\begin{document}

\title{Macroscopic Expression Connecting the Rate of Energy Dissipation and Violation of the Fluctuation-Response Relation}

\author{Takahiro Harada}
\email[Electronic address: ]{harada@phys.s.u-tokyo.ac.jp}
\affiliation{Department of Physics, Graduate School of Science, the University of Tokyo, Tokyo 113-0033, Japan}
\affiliation{Department of Human and Artificial Intelligent Systems, University of Fukui, Fukui 910-8507, Japan}

\date{\today}

\begin{abstract}
A direct connection between the magnitude of the violation of the fluctuation-response relation (FRR) and the rate of energy dissipation is presented in terms of field variables of nonequilibrium systems.
Here, we consider the density field of a colloidal suspension either in a relaxation process or in a nonequilibrium steady state driven by an external field.
Using a path-integral representation of the temporal evolution of the density field, we find an equality that relates the magnitude of the violation of the FRR for scalar and vector potentials of the velocity field to the rate of energy dissipation for the entire system.
Our result demonstrates that the violation of the FRR for field variables captures the entropic component of the dissipated free energy.
\end{abstract}


\maketitle

In recent years, considerable effort has been made to understand the nature of fluctuations in nonequilibrium systems.
Of particular note are studies on the fluctuation theorems \cite{Gallavotti:1995, Crooks:1997, Maes:1999}, the Jarzynski equality \cite{Jarzynski:1997}, the Hatano-Sasa equality \cite{Hatano:2001}, and the other related equalities that hold exactly for systems driven far from equilibrium.
In addition to these works on rather established topics, further novel results regarding the properties of nonequilibrium systems are still being obtained \cite{Komatsu:2008a, Komatsu:2008b}.

With regard to fluctuations in nonequilibrium systems, we recently reported on the physical significance of the violation of the fluctuation-response relation (FRR) in systems far from equilibrium \cite{Harada:2005, Harada:2006}.
The FRR is a relation between the response function and the correlation function of gross variables.
This relation has been firmly established for systems near equilibrium \cite{Kubo:1991}, and it is known to be violated in various nonequilibrium systems \cite{Cugliandolo:1997, Crisanti:2003, Harada:2005, Harada:2006}. In previous works \cite{Harada:2005, Harada:2006}, we presented several equalities that directly relate the amount by which the FRR is violated and the rate of energy dissipation for systems described by Langevin equations.
These equalities have been extended in several ways \cite{Deutsch:2006, Saito:2008} and have been experimentally verified \cite{Toyabe:2007, Blickle:2007, Toyabe:2008}.
In these studies, violation of the FRR is considered in terms of mechanical degrees of freedom.
It has not yet been demonstrated whether a similar equality also holds for field variables, e.g., the density field of the particles.
In the present Rapid Communication, we demonstrate that in fact violation of the FRR for field variables is also directly connected to the dissipation rate.
Our result provides new insight into violation of the FRR in nonequilibrium systems as well as the proposal of a novel method to experimentally probe energy dissipation in nonequilibrium soft materials.

For the sake of concreteness, here we consider $n$ spherical colloidal particles suspended in an aqueous solution of temperature $T$ and confined to a three-dimensional box \cite{footnote:extension}.
The particles are subject to a potential field, $U(\vec x)$, and a static driving force, $\vec F(\vec x)$. We suppose that $\vec F(\vec x)$ cannot be written in a gradient form, and it thus represents a nonconservative force.
The particles are interacting with each other via a potential represented by $W(\vec x)$. For simplicity, we assume $W(\vec x) = W(-\vec x)$.
For such a system, the motion of the $i$-th particle is described by the following Langevin equation:
\begin{equation}
\gamma \dot {\vec x}_i = \vec F(\vec x_i) - \nabla U(\vec x_i) - \sum_{j=1}^n \nabla W(\vec x_i - \vec x_j) + \vec\xi_i.
\label{e.model}
\end{equation}
Here, $\gamma$ denotes the friction coefficient of a particle, and $\vec \xi_i(t)$ represents zero-mean white Gaussian noise satisfying $\bra \vec \xi_i (t) \vec \xi_j (s)\ket = 2\gamma T \delta_{ij} \tens{I} \delta (t-s)$, where $\tens{I}$ is unit matrix.
The Boltzmann constant is set to unity.
For simplicity, we do not consider the hydrodynamic interactions among particles in the present Rapid Communication.
Note that the model (\ref{e.model}) has been considered in many contexts, including the rheology of colloidal suspensions \cite{Fuchs:2005}, glass transitions \cite{Kawasaki:1994}, and fundamental studies of nonequilibrium steady states \cite{Nakamura:2006}.

Here, we are interested in fluctuations and the response of the density field of the particles denoted by $\rho(\vec x, t)$.
In order to study the response of the system, one can introduce a perturbative external force, which we write as $\vec f(\vec x_i, t) = -\ve \nabla \phi(\vec x_i, t) + \lambda \nabla \times \vec A(\vec x_i, t)$ ($\ve, \lambda \ll 1$), and add this to the r.h.s.~of (\ref{e.model}).
In particular, two types of responses are important in the following.
One is the response of a scalar field, $X[[\rho]] (\vec x, t)$, which is a functional of $\rho$, to the scalar field $\ve \phi$, written in the form
\begin{eqnarray}
\lefteqn{\bra X (\vec x, t) \ket_{\ve, 0} = \bra X (\vec x, t) \ket_0} \nonumber \\
& & - \ve \int_0^t d t' \int_\Omega d \vec x R_X (\vec x, t; \vec x', t') \phi (\vec x', t') + O(\ve^2),
\label{e.R_def}
\end{eqnarray}
which defines a response function $R_X$. Here, $\bra \cdot \ket_{\ve, \lambda}$ ($\bra \cdot \ket_0$) represents the ensemble average in the presence (absence) of the perturbation fields, and $\Omega$ denotes a sufficiently large region that contains the box. We suppose that $\phi$ and $\vec A$ vanish at the boundary of $\Omega$.
The other type of response is that of a vector field, $\vec Y[[\rho]] (\vec x, t)$, which is also a functional of $\rho$, to the vector potential $\lambda \vec A$. This response is written in the form
\begin{eqnarray}
\lefteqn{\bra \vec Y (\vec x, t) \ket_{0, \lambda} = \bra \vec Y (\vec x, t) \ket_0} \nonumber \\
& & - \lambda \int_0^t d t' \int_\Omega d \vec x \tens Q_{\svec Y} (\vec x, t; \vec x', t') \cdot \vec A(\vec x', t') + O(\lambda^2), \hspace{5mm}
\label{e.Q_def}
\end{eqnarray}
which defines a response matrix $\tens Q_{\svec Y}$.
Note that $R_X (\vec x, t; \vec x', t')$ and $\tens Q_{\svec Y}(\vec x, t; \vec x', t')$ are defined only for $t \ge t'$, while we set them to zero for $t < t'$ , in accordance with causality.

Next, fluctuations of the density field can be characterized by the correlation function.
Representing scalar functionals of $\rho$ by $X[[\rho]] (\vec x, t)$ and $X'[[\rho]] (\vec x', t')$, the correlation function is defined as
\[
C_{X, X'}(\vec x, t; \vec x', t') \equiv \bra X(\vec x, t) X'(\vec x', t') \ket_0.
\]
Then, for vector functionals of $\rho$, denoted by $\vec Y[[\rho]] (\vec x, t)$ and $\vec Y'[[\rho]] (\vec x', t')$, we define
\[
\tens D_{\svec Y, \svec Y'} (\vec x, t; \vec x', t') \equiv \bra \vec Y(\vec x, t) \vec Y'(\vec x', t') \ket_0
\]
as the correlation matrix.

In the present Rapid Communication, we show that the quantities defined above are directly connected to the total rate of free energy dissipation in the system, defined as
\begin{equation}
J_\Omega(t) \equiv - \der{\F[\rho(\cdot, t)]}{t} + \int_\Omega d\vec x \vec j(\vec x, t) \cdot \vec F(\vec x),
\label{e.J}
\end{equation}
where $\F[\rho(\cdot, t)]$ represents the free energy of the system defined in (\ref{e.energy}) and $\vec j(\vec x, t)$ is the mass flux defined in (\ref{e.continuity}). Our central result is the following \cite{footnote:divergence}:
\begin{eqnarray}
\lefteqn{\bra J_\Omega (t) \ket_0 = \gamma \int_\Omega d \vec x  \Big\{ C_{\V, -\nabla \cdot \svec j}(\vec x, t; \vec x, t) - T R_\V (\vec x, t; \vec x, t) } \nonumber \\
& & + \tr\left[ \tens D_{\Ws, - \nabla \times \svec j} (\vec x, t; \vec x, t) - T \tens Q_{\Ws}(\vec x, t; \vec x, t) \right] \Big\}. \hspace{10 mm}
\label{e.central}
\end{eqnarray}
Here, $\V$ and $\W$ represent scalar and vector potentials of the velocity field, $\vec j(\vec x, t)/\rho(\vec x, t)$, defined as
\begin{eqnarray}
\V [[\rho]] (\vec x, t) &\equiv& - \int_\Omega d \vec x' g(\vec x; \vec x') \nabla' \cdot \left( \frac{\vec j(\vec x', t)}{\rho(\vec x', t)} \right), \label{e.v_def} \\
\W [[\rho]] (\vec x, t) &\equiv& - \int_\Omega d \vec x' g(\vec x; \vec x') \nabla' \times \left( \frac{\vec j(\vec x', t)}{\rho(\vec x', t)} \right), \hspace{5mm}
\label{e.w_def}
\end{eqnarray}
where $g(\vec x; \vec x')$ is the Green function of the Laplacian operator in $\Omega$, defined as
$
\Delta g(\vec x; \vec x') = - \delta(\vec x - \vec x'),
$
with the boundary condition $g(\vec x; \vec x') = 0$ for $\vec x \in \partial \Omega$.

Interestingly, the r.h.s.~of (\ref{e.central}) can be interpreted as the magnitude of the violation of the following FRRs:
\begin{eqnarray}
C_{\V, -\nabla \cdot \svec j}(\vec x, t; \vec x', t') &=& T R_\V (\vec x, t; \vec x', t'), \label{e.FRR_R}\\
\tens D_{\Ws, - \nabla \times \svec j} (\vec x, t; \vec x', t') &=& T \tens Q_{\Ws}(\vec x, t; \vec x', t'),
\label{e.FRR_Q}
\end{eqnarray}
As demonstrated below, (\ref{e.FRR_R}) and (\ref{e.FRR_Q}) hold when the system is in equilibrium.
However, it has been argued that when the system is far from equilibrium, the FRRs (\ref{e.FRR_R}) and (\ref{e.FRR_Q}) do not hold in general \cite{Cugliandolo:1997, Crisanti:2003, Harada:2005, Harada:2006}.
The equality (\ref{e.central}) demonstrates that there is a direct relation between the rate of energy dissipation and the magnitude of the violation of the FRR for $\V$ and $\W$ for a system far from equilibrium.

It is also noteworthy that the r.h.s.~of the equality (\ref{e.central}) consists entirely of experimentally measurable quantities.
Thus, using this equation, we can investigate the energetic aspects of a colloidal suspension driven far from equilibrium without the need for a detailed study of the force fields such as $\vec F, U$, and $W$.
In fact, all the quantities on the r.h.s.~of (\ref{e.central}), except for $\gamma$, can be determined through measurements of the density field and the flux by considering cases with and without the perturbation fields.
For a colloidal suspension, it might be possible to make such measurements by using a confocal microscopy \cite{Weeks:2000}, for example, and precisely calibrated external fields can be created using a variety of techniques, such as those employing optical trap arrays \cite{Korda:2002}.
Furthermore, the value of $\gamma$ can be estimated with various well-established methods, e.g., using the Einstein-Stokes relation or through measurement of the high-frequency response \cite{Harada:2006}.

We now derive (\ref{e.central}).
We start with the fluctuating hydrodynamic equation describing the time evolution of the density field, $\rho(\vec x, t)$, which has been obtained by several authors \cite{Kawasaki:1994, Dean:1996, Andreanov:2006, Nakamura:2009}:
\begin{eqnarray}
\pder{}{t} \rho(\vec x, t) &=& - \nabla \cdot \vec j (\vec x, t), \label{e.continuity} \\
\vec j(\vec x, t) &=& \frac{\rho(\vec x, t)}{\gamma}  \Big[ \vec F(\vec x) - \nabla \fder{\F [\rho(\cdot, t)]}{\rho(\vec x, t)} -\ve\nabla \phi(\vec x, t) \nonumber \\
& & + \lambda \nabla \times \vec A(\vec x, t) \Big] + \frac{\sqrt{\rho(\vec x, t)}}{\gamma}  \circ \vec \zeta(\vec x, t), \label{e.flux}
\end{eqnarray}
where $\F$ represents the free energy functional, defined as
\begin{eqnarray}
\F[\vphi] &\equiv& \int_\Omega d\vec x U(\vec x)\vphi(\vec x) + T\int_\Omega d \vec x \vphi(\vec x) \left[ \ln \vphi(\vec x) - 1 \right] \nonumber \\
& & + \frac{1}{2} \int_\Omega d \vec x \int_\Omega d \vec y \vphi(\vec x) W(\vec x - \vec y) \vphi(\vec y),
\label{e.energy}
\end{eqnarray}
and $\vec \zeta (\vec x, t)$ represents white Gaussian noise satisfying $\bra \vec \zeta(\vec x, t) \ket = 0$ and $\bra \vec \zeta(\vec x, t) \vec \zeta(\vec y, s) \ket = 2\gamma T \tens{I} \delta(\vec x - \vec y) \delta(t - s)$. The multiplication $\circ$ is interpreted according to the It\^o rule for both space and time \cite{Gardiner:2004}.
Next, using the Green function, $g(\vec x; \vec x')$, and the Helmholtz theorem, (\ref{e.continuity}) and (\ref{e.flux}) can be rearranged as
\begin{eqnarray}
\vec \eta(\vec x, t) &=& \nabla \fder{\F [\rho(\cdot, t)]}{\rho(\vec x, t)} - \vec F(\vec x) + \frac{\gamma}{\rho(\vec x, t)} \nabla \times  \vec \sigma (\vec x, t) \nonumber \\
& & + \frac{\gamma}{\rho(\vec x, t)} \int_\Omega d \vec x' \nabla g(\vec x; \vec x') \pder{}{t} \rho(\vec x', t) \nonumber \\
& & + \ve \nabla \phi (\vec x, t) - \lambda \nabla \times \vec A(\vec x, t),
\label{e.eta_def}
\end{eqnarray}
where $\vec \eta(\vec x, t) \equiv \rho(\vec x, t)^{-1/2} \circ \vec \zeta(\vec x, t)$, and $\vec \sigma (\vec x, t)$ is a vector field satisfying $\nabla \cdot \vec \sigma (\vec x, t) = 0$. 
Thus, $\vec \sigma (\vec x, t)$ can be specified by two independent fields, which we write $\sigma_1 (\vec x, t)$ and $ \sigma_2 (\vec x, t)$.
Let $\vec \rho$ denote $(\rho, \sigma_1, \sigma_2)$ collectively.
It can be shown that the noise $\vec \eta(\vec x, t)$ is zero-mean white Gaussian noise satisfying $\bra \vec \eta (\vec x, t) \vec \eta (\vec x', t') \ket = 2\gamma T \tens{I} \delta(\vec x - \vec x')\delta(t - t')/\rho(\vec x, t)$.
The relation (\ref{e.eta_def}) defines a transformation between $\vec \eta(\vec x, t)$ and $\vec \rho(\vec x, t)$. Note that the functional derivative $\delta \vec \eta/\delta \vec \rho$ is independent of $\ve \phi$ and $\lambda \vec A$, because these fields are independent of $\vec \rho$. Therefore, the Jacobian of the transformation, $J[[\vec \rho]] \equiv \det(\delta \vec \eta/\delta \vec \rho)$, is also independent of the perturbative fields.
We thus obtain the path-integral representation for the transition probability of $\vec \rho$ from $t = 0$ to $t = \tau$ under a given initial condition, $\vec \rho_0 \equiv \vec \rho(\cdot, 0)$, as
\begin{equation}
P_{\ve, \lambda} ([[\vec \rho]] | [\vec \rho_0]) = J[[\vec \rho]] N[[\vec \rho]] \exp(-\beta S_{\ve, \lambda} [[\vec \rho]]),
\end{equation}
where $N[[\vec \rho]] \equiv \prod_{\svec x} \prod_t [(\rho(\vec x, t)/4\pi \gamma T)^3 d\vec x d t]^{1/2}$ is a normalization constant for the path probability of $\vec \eta$, and $\beta \equiv T^{-1}$. Note that $N[[\vec \rho]]$ is also independent of the perturbation fields. The action, $S_{\ve, \lambda} [[\vec \rho]]$, is given by
\begin{eqnarray*}
\lefteqn{ S_{\ve, \lambda} [[\vec \rho]] = \frac{1}{4\gamma} \int_0^\tau d t \int_\Omega d \vec x \rho(\vec x, t) \Big[ \nabla \fder{\F [\rho(\cdot, t)]}{\rho(\vec x, t)} - \vec F(\vec x)}\nonumber \\
& & + \frac{\gamma}{\rho(\vec x, t)} \left( \nabla \times \vec \sigma (\vec x, t) + \int_\Omega d \vec x' \nabla g(\vec x; \vec x') \pder{}{t} \rho(\vec x', t) \right) \nonumber \\
& & +\ve \nabla \phi(\vec x, t) - \lambda \nabla \times \vec A(\vec x, t) \Big]^2.
\end{eqnarray*}

Our next task is the calculation of the average of the functionals $X[[\rho]](\vec x, t)$ and $\vec Y[[\rho]] (\vec x, t)$.
The average of $X$ is defined as
\[
\bra X (\vec x, t) \ket_{\ve, \lambda} \equiv \int \D \vec \rho X [[\rho]] (\vec x, t) P_{\rm i}[\vec \rho_0] P_{\ve, \lambda} ([[\vec \rho]] | [\vec \rho_0]).
\]
The initial distribution for the density fields, $P_{\rm i} [\vec \rho_0]$, can be chosen arbitrarily at this point.
We then study separately the effects of the two types of perturbations, $\ve \phi$ and $\lambda \vec A$.
First, we consider the former by setting $\lambda =0$.
After setting $t = \tau/2$, a straightforward calculation leads to
\begin{eqnarray}
\lefteqn{ \bra X (\vec x, \tau/2) \ket_{\ve, 0} = \bra X (\vec x, \tau/2) \ket_0 } \nonumber \\
& & - \ve \int_0^\tau d t' \int_\Omega d\vec x' \frac{\beta}{2\gamma} \big[  C_{X, \nabla \cdot \rho \svec G} (\vec x, \tau/2; \vec x', t') \nonumber \\
& & + \gamma C_{X, -\nabla \cdot \svec j}(\vec x, \tau/2; \vec x', t') \big] \phi(\vec x', t') + O(\ve^2),
\label{e.X_mean}
\end{eqnarray}
where $\vec G(\vec x, t) \equiv \vec F(\vec x) - \nabla \delta \F[\rho(\cdot, t)]/\delta \rho(\vec x, t)$ and the fact that $\rho(\vec x, t) = 0$ for $\vec x \in \partial \Omega$, which follows because the particles are confined to the box, has been utilized.
We then restrict the perturbation field so as to satisfy $\phi(\vec x, \tau - t) = \phi(\vec x, t)$. This allows us to compare (\ref{e.X_mean}) with (\ref{e.R_def}), and we thereby obtain the following expression:
\begin{eqnarray}
\lefteqn{ T R_X (\vec x, \tau/2; \vec x', t') = \tilde C_{X, -\nabla \cdot \svec j}(\vec x, \tau/2; \vec x', t') } \hspace{20mm} \nonumber \\
& & + \tilde C_{X, \nabla \cdot \rho \svec G} (\vec x, \tau/2; \vec x', t')/\gamma,
\label{e.R_general}
\end{eqnarray}
where tilde denotes the time-reversal symmetric part, defined as, e.g., $\tilde B(\vec x, t; \vec x', t') \equiv (B(\vec x, t; \vec x', t') + B(\vec x, \tau - t; \vec x', \tau - t'))/2$.
Next, setting $\ve = 0$, a similar calculation for $\bra \vec Y (\vec x, t) \ket_{0, \lambda}$ under the assumption $\vec A(\vec x, t) = \vec A(\vec x, \tau - t)$ yields the following expression for $\tens Q_{\svec Y} (\vec x, t; \vec x', t')$:
\begin{eqnarray}
\lefteqn{T \tens Q_{\svec Y} (\vec x, \tau/2; \vec x', t') = \tilde {\tens D}_{\svec Y, -\nabla \times \svec j}(\vec x, \tau/2; \vec x', t')} \hspace{20mm} \nonumber \\
& & + \tilde {\tens D}_{\svec Y, \nabla \times \rho \svec G} (\vec x, \tau/2; \vec x', t')/\gamma. \label{e.Q_general}
\end{eqnarray}
Here, the equality $\Delta \vec \sigma (\vec x, t) = - \nabla \times \vec j(\vec x, t)$, which follows from (\ref{e.flux}) and (\ref{e.eta_def}), has been used.
Equations (\ref{e.R_general}) and (\ref{e.Q_general}) are valid for arbitrary functionals, $X$ and $\vec Y$, and they express the relations among the response functions and correlation functions.

We can obtain several important relations from (\ref{e.R_general}) and (\ref{e.Q_general}).
First, the FRRs (\ref{e.FRR_R}) and (\ref{e.FRR_Q}) can be obtained by setting $\vec F(\vec x) = \vec 0$ and preparing the system with the canonical distribution, 
$
P_{\rm c} [\vec \vphi] = Z^{-1} \exp(-\beta \F[\vphi])
$
with $Z \equiv \int \D \vec \vphi \exp(-\beta \F[\vphi])$. In this case, if $X$ and $\vec Y$ satisfy time-reversal antisymmetry, i.e., $X[[\hat {\vec \rho}]] = -X[[\vec \rho]]$ and $\vec Y[[\hat {\vec \rho}]] = - \vec Y[[\vec \rho]]$, where $\hat{\vec \rho} (\vec x, t) \equiv \vec \rho(\vec x, \tau - t)$, it can be confirmed that the second terms on the r.h.s.~of (\ref{e.R_general}) and (\ref{e.Q_general}) vanish, as a consequence of the detailed balance condition
$
P_0([[\vec \rho]] | [\vec \rho_0]) P_{\rm c}[\vec \rho_0] = P_0([[\hat{\vec \rho}]] | [\vec \rho_\tau]) P_{\rm c}[\vec \rho_\tau],
$
where $\vec \rho_\tau \equiv \vec \rho(\cdot, \tau)$.
Because $\V[[\rho]](\vec x, t)$ and $\W[[\rho]](\vec x, t)$ satisfy time-reversal antisymmetry at $t = \tau/2$, we obtain (\ref{e.FRR_R}) and (\ref{e.FRR_Q}) from (\ref{e.R_general}) and (\ref{e.Q_general}) by setting $X = \V$, $\vec Y = \W$ and $t = \tau/2$.

In contrast to the case analyzed above, when $\vec F(\vec x) \neq \vec 0$ and/or $P_{\rm i} \neq P_{\rm c}$, the system is not in equilibrium.
In this case, we can obtain (\ref{e.central}) from (\ref{e.R_general}) and (\ref{e.Q_general}) as follows.
First, if we set $\vec x' = \vec x$ and $t' = \tau/2$, (\ref{e.R_general}) and (\ref{e.Q_general}) yield
\begin{eqnarray}
\lefteqn{ \gamma \left[ C_{X, -\nabla \cdot \svec j}(\vec x, t; \vec x,t) - T R_X (\vec x, t; \vec x, t) \right] =} \nonumber \\ 
& & \hspace{20mm} - \bra X(\vec x,t) \nabla \cdot \left[ \rho(\vec x, t) \vec G(\vec x, t) \right] \ket_0, \label{e.central_R} \\
\lefteqn{ \gamma \tr \left[ \tens D_{\svec Y, -\nabla \times \svec j}(\vec x, t; \vec x,t) - T \tens Q_{\svec Y} (\vec x, t; \vec x, t) \right] =} \nonumber \\ 
& & \hspace{20mm} - \bra \vec Y(\vec x,t) \cdot \left[ \nabla \times \rho(\vec x, t) \vec G(\vec x, t) \right] \ket_0, \hspace{5mm}
\label{e.central_Q}
\end{eqnarray}
where $t'$ has been replaced with $t$.
We then integrate the sum of (\ref{e.central_R}) and (\ref{e.central_Q}) over $\Omega$ to obtain
\begin{eqnarray}
\lefteqn{\gamma \int_\Omega d\vec x \big\{ C_{X, -\nabla \cdot \svec j}(\vec x, t; \vec x,t) - T R_X (\vec x, t; \vec x, t) } \nonumber \\
& & + \tr \left[ \tens D_{\svec Y, -\nabla \times \svec j}(\vec x, t; \vec x,t) - T \tens Q_{\svec Y} (\vec x, t; \vec x, t) \right] \big\} = \nonumber \\
& & \hspace{15mm} \Big\langle \int_\Omega d\vec x \rho(\vec x, t) \vec V(\vec x, t) \cdot \vec G(\vec x, t) \Big\rangle_0,
\label{e.balance}
\end{eqnarray}
where $\vec V \equiv \nabla X - \nabla \times \vec Y$ and the fact that $\rho(\vec x, t) = 0$ for $\vec x \in \partial \Omega$ has again been utilized.
We next set $X = \V$ and $\vec Y = \W$. Because the relation $\nabla \V(\vec x, t) - \nabla \times \W(\vec x, t) = \vec j(\vec x, t)/\rho(\vec x, t)$ follows from (\ref{e.v_def}) and (\ref{e.w_def}), the integral on the r.h.s.~of (\ref{e.balance}) can be rewritten as
\begin{eqnarray}
\int_\Omega d\vec x \vec j(\vec x, t) \cdot \vec G(\vec x, t) = J_\Omega (t),
\label{e.dissipation}
\end{eqnarray}
where we have used the fact that $\vec j(\vec x, t) = \vec 0$ for $\vec x \in \partial \Omega$.
Finally, combining (\ref{e.balance}) and (\ref{e.dissipation}), we obtain the desired expression given in (\ref{e.central}).

In this way, we have obtained a concise relation between the rate of energy dissipation and the magnitude of the FRR violation for field variables.  Before closing this Rapid Communication, we make several remarks on the significance of this result.

A model similar to (\ref{e.model}) has also been studied in our previous report \cite{Harada:2006}.
There, an equality relating the dissipation rate and the magnitude of the violation of the FRR for each mechanical degree of freedom was derived. The present result (\ref{e.central}) contains the previous result in Ref.~\cite{Harada:2006} as a special case.
If the spatial resolution of our measurement is sufficiently good so that we can distinguish every particle, the density field can be defined using the coordinates of the particles as $\rho(\vec x, t) \equiv \sum_{i=1}^n \delta(\vec x - \vec x_i(t))$. In this case, the entropy in (\ref{e.energy}), i.e., $- \int_\Omega d \vec x \rho(\vec x, t) [\ln \rho(\vec x, t) - 1]$, becomes independent of the configuration of the particles due to the delta-function nature of the density field. Therefore, the contribution of the entropy term to the total dissipation (\ref{e.J}) vanishes, and the dissipation arises from the purely mechanical origin. In this situation, we can prove the mathematical equivalence between the present result (\ref{e.central}) and the previous result in \cite{Harada:2006} (see \cite{Harada:inprep, Yoshimori:priv}). In contrast, if our spatial resolution is restricted and we cannot distinguish each particle, the entropy changes in time and it measures the loss of information on the arrangement of the particles. In this case, the variation of the entropy term also contributes to the dissipation (\ref{e.J}). This is in accordance with Sekimoto's discussion that the definition of dissipation should change depending on the level of description \cite{Sekimoto:2007}.
Our analysis here remains valid even in this case, and the FRR violation in the r.h.s.~of (\ref{e.central}) can capture the entropic component of the dissipation.

The present result in (\ref{e.central}) has also a practical advantage compared to the previous result in \cite{Harada:2006}.
In our previous result, in order to measure the FRR violation, one has to track each particle and to apply spatially uniform perturbative force separately to every particle.  This is quite difficult to be experimentally achieved. In contrast, measurement of the FRR violation in the present equality (\ref{e.central}) does not require the identification of each particle. Moreover, we can utilize spatially nonuniform fields, $\phi$ and $\vec A$, that can be applied to the entire system at one time. The latter should be more easily done in experiments.
Therefore, the present approach extends the possibility to study the energetic aspect of nonequilibrium systems through measurement of the FRR violation.

\begin{acknowledgments}
The author acknowledges suggestions by T.~Ohta and T.~Yamaguchi, which provided the initial motivation for this study, valuable communications with S.~-i.~Sasa and A.~Yoshimori, and conversations with M.~Miyazaki.
This work was partly supported by a grant-in-aid from JSPS (No.~20740239).
\end{acknowledgments}


\begin{thebibliography}{99}
\bibitem{footnote:extension} The analysis presented here can be extended to systems of other dimensions with proper modifications \cite{Harada:inprep}.
\end{thebibliography}
\end{document}